\documentclass[a4paper,twoside]{article}
\newcommand{\affil}[1]{$^{\rm #1}$}
\usepackage[authoryear]{natbib}
\bibpunct{(}{)}{;}{a}{}{,}
\usepackage{graphicx}
\date{} 
\newcommand{\Hband}{\textit{H}-band\ }
\title{\large\bf\flushleft The Near-Infrared Surface Brightness
  Distribution of NGC\,4696}
\author{\parbox{\textwidth}{\flushleft
\vspace{-0.5cm}
{\it P.\,Arnalte Mur\affil{A,B,C}, S.\,C.\,Ellis\affil{A}, 
     and Matthew Colless\affil{A}}\\
\vspace{0.4cm}
{\small \affil{A}\,Anglo-Australian Observatory, P.O. Box 296, 
                   Epping, NSW 2121, Australia}\\
{\small \affil{B}\,Facultat de F\'{i}sica, Universitat de Val\`{e}ncia, 
                   46100 Burjassot, Val\`{e}ncia, Spain}\\
{\small \affil{C}\,Email: parmur@alumni.uv.es}}}
\begin{document}
\twocolumn[
\begin{changemargin}{.8cm}{.5cm}
\begin{minipage}{.9\textwidth}
\vspace{-1cm}
\maketitle
%
%
\small{\bf Abstract:}

We present \Hband observations of the elliptical galaxy NGC\,4696, the
brightest member of the Centaurus cluster of galaxies. We have measured
its light profile, using a two-dimensional fitting algorithm, out to a
radius of 180\,arcsec ($37\,h^{-1}_{70}\,\mathrm{kpc}$). The profile is
well described by a de~Vaucouleurs law, with an effective radius of
$35.3\pm1.0\,h^{-1}_{70}\,\mathrm{kpc}$. There is no need for the extra
free parameter allowed by a S\'{e}rsic law. Allowing for a variation of
0.3\% in the sky level, the profile obtained is compatible with data
from 2MASS. The profile shows no sign of either a truncation or an
extended halo.


\medskip
\medskip
\end{minipage}
\end{changemargin}
]
\small

\section{Introduction}
\label{sec:intro}

Surface brightness profiles of elliptical galaxies are normally well
described by either a de~Vaucouleurs law \citep{devaucouleurs48} or a
S\'{e}rsic law \citep{sersic68}. However, the situation can be more
complicated when we consider a galaxy embedded in a cluster. In this
case, the outer parts of the profile may be modified due to tidal
interactions with neighbouring galaxies, giving a `truncated' profile. A
different case is often encountered in elliptical galaxies that are the
brightest cluster member (BCM) and are located at the cluster centre,
many of which have an extended envelope or halo
\citep{schombert87,gonzalez05}. These galaxies are known as `central
dominant' (cD) galaxies. The extended envelope is observed as a
deviation of the profile from a de~Vaucouleurs or S\'{e}rsic law at
large radii, and appears to correspond to light from stars orbiting in
the cluster potential and not bound to the central galaxy
\citep*{bahcall77, gonzalez05}. Simulations indicate that the stars in
the envelope may, on average, be older than the stars in the galaxy
\citep{murante04}.

NGC\,4696 is the BCM of the Centaurus cluster. It is an elliptical
galaxy located near the centre of the cluster. It is sometimes described
as a cD galaxy, although no evidence has been found up to now of an
extended halo. 

Previous measurements of the light profile of NGC\,4696 at optical
wavelengths include those of \citet{schombert87} and \citet{jerjen97},
both obtained from photographic plates. \citet{jerjen97} measured the
light profile out to a radius of $\approx$80\,arcsec
($\approx16\,h^{-1}_{70}\,\mathrm{kpc}$), while the profile obtained by
\citet{schombert87} reaches much further, out to a radius of
$\approx$9\,arcmin ($\approx110\,h^{-1}_{70}\,\mathrm{kpc}$). Neither
observation showed signs of an extended envelope. The profile from
\citet{schombert87} followed an $r^{\frac{1}{4}}$ law (de~Vaucouleurs
profile) out to a radius of $\approx$5\,arcmin, and a truncation or
`distention' of the profile was found for larger radii.

Measurements of the NIR profile were obtained by the Two Micron All Sky
Survey (2MASS source 2MASX~J12484927-4118399 in the All-Sky Extended
Source Catalog; http://www.ipac.caltech.edu/2mass/). Light profiles were
obtained in three bands (\textit{J}, \textit{H} and \textit{K}) for
radii up to $\approx$3\,arcmin. This corresponds to surface brightnesses
$\mu_{H} \leq 24$\,mag\,arcsec$^{-2}$ in the \textit{H} band.

The Intra-Cluster Medium (ICM) of the Centaurus cluster has been studied
using X-ray emission by the \textit{ROSAT} and \textit{ASCA} satellites.
\citet{allen94}, using data from \textit{ROSAT}, found a correlation
between the structure---ellipticity and centroid---of the X-ray emission
and optical observations \citep{sparks89} of the central galaxy
NGC\,4696. \citet{ikebe99} used data from both satellites, and described
the X-ray emitting gas in the central region ($r \leq
3\,\mathrm{arcmin}$) according to a two-phase model. They identified the
cool phase with the inter-stellar medium (ISM) of NGC\,4696, and the hot
phase with the ICM of the Centaurus cluster. These relationships between
the central galaxy and the ICM could be an indication of the presence of
an extended halo.

The previous optical and NIR work pose a quandary in that the form of the profiles measured range from de Vaucouleurs to exponential.  Thus the aim of the present work was to observe the light
profile of NGC\,4696 in the near-infrared (NIR) out to large radii,
study the functional form of the profile (de Vaucouleurs or S\'{e}rsic
law), measure the parameters of the profile, and seek evidence for
either a truncation due to tidal interactions or an extended envelope
following the cluster potential.  The main advantage of the NIR  over optical studies is that the light is dominated by old stars, and thus the NIR profile will better trace the bulk of the stellar mass of the galaxy.  Furthermore the extinction due to dust is smaller in the NIR than at optical wavelengths.  

Another motivation for this work is the application to studying radial colour distributions, through coupling with optical data, to understand the distribution of stellar populations throughout BCMs.  Through comparisons with simulations (e.g.\ \citealt{som05}), such studies would help to constrain BCM and cluster formation theories and the enrichment of the intra-cluster medium (e.g.\ \citealt{lin04}).

We describe our observational data in Section~\ref{sec:observ} and
explain the reductions and calibrations in Section~\ref{sec:red} and the
method used to fit the profile models in Section~\ref{sec:an}.
Section~\ref{sec:results} presents the results and
Section~\ref{sec:concl} summarises our conclusions. 

Throughout the paper we use a Hubble constant of
$H_{0}=70\,h_{70}$\,km\,s$^{-1}$\,Mpc$^{-1}$. We have used a redshift of $z=0.00987\pm0.00005$ (\citealt{devaucouleurs91}, which assumes no peculiar velocity) which equates to a distance of
$42.3\pm0.2\,h_{70}^{-1}$\,Mpc; the angular scale at this distance is
$0.205\pm0.001$\,kpc\,arcsec$^{-1}$.  Distances derived from surface brightness fluctuations, taking into account the peculiar motions of the Cen30 and Cen45 components of the Centaurus cluster, and the Hydra cluster, give a distance of $42.5 \pm 3.2$ Mpc, consistent with the value used here (\citealt{mie05}).

\section{Observations}
\label{sec:observ}

Observations were made on 2005 February 16, using the AAO Infrared
Imager \& Spectrograph (IRIS2) at the 3.9m Anglo-Australian Telescope
(AAT). Conditions during the night were photometric, but the seeing was
2.5\,arcsec. Data were obtained using the \textit{H} filter. The detector
in IRIS2 is a 1024$\times$1024 Rockwell HAWAII-1 HgCdTe infrared
detector. The detector scale is 0.4486\,arcsec\,pixel$^{-1}$, giving a
field of view of 7.7$\times$7.7\,arcmin \citep{tinney04}.

We obtained a total of 99 frames of the field around NGC\,4696 using a
9-point dither. Each of these frames is an average of 6~exposures of 10s
each. To measure the light profile to the full extent allowed by the
detector, NGC\,4696 was placed in the south-east corner of the field of
view. We also obtained observations of three photometric standard stars
(UKIRT Faint Standards FS129, FS132 and FS135; \citealt{hawarden01}),
taking five 10s exposures in each case. In order to model the point
spread function (PSF), a bright star at a similar airmass to the science
field was observed over 93 2.5s exposures. As explained in
Section~\ref{sec:red}, extra data was needed for the flat-fielding
process. This was obtained from other \Hband observations taken on
February 11 during the same run.

\section{Reduction and Calibration}
\label{sec:red}

The data reduction was carried out using the IRAF (http://iraf.noao.edu/)
and ORAC-DR (http://www.oracdr.org/) software. The steps followed were:
subtraction of dark current, flat-fielding, registering to obtain the
correct astrometric offsets, and combination of the 99 frames.

In the first step, the dark current in each pixel was subtracted from
each of the raw frames. This value was obtained from a dark frame taken
using the same exposure time (10s) as the science frames. The typical
value of the dark current was $\sim$3\,ADU ($\sim$15\,electrons).

To take into account the difference in response between pixels in the
detector, it was necessary to flat-field the frames. It was not possible
to obtain a flat field directly from our science field, as the large
extent of NGC\,4696 meant that any flat field obtained from the science
field images would retain some contribution from the galaxy.

The flat field used was therefore derived from \Hband observations made
on February 11, during the same IRIS2 observing run. These were used
instead of other observations from February 16 for a variety of reasons.
In the case of the standard photometric stars, other very bright stars
in the field caused artifacts in the flat field. In the case of the `PSF
star', the accuracy of the resulting flat field was limited by the short
exposure time. Using all the 10s exposed frames from February 11 that
did not contain bright stars, we could construct a flat field with a
total exposure time of about 58\,min.

The flat field from these frames was obtained using ORAC-DR. For each of
the fields, the frames were median-combined to obtain a first estimate
for the flat. Then the original frames were divided by this estimate,
objects were detected and masked, and the frames were combined again to
obtain the final flat for the field. The flat used was an average of
these flats. The dark-subtracted frames were divided by this flat before
being registered and combined.

Before combining the 99 frames, registration was carried out to find the
true offsets between the frames. This step was done using the 
{\tt xregister} task in IRAF, which performs a cross-correlation between
objects in the field. Knowing these offsets, the IRAF task {\tt combine}
was used to sigma-clip and median-combine the frames, obtaining the
final image. The base sky-level is not subtracted, but rather an additive zero point is applied to each frame  to account
for differences in sky level. This zero point was calculated as the mode
of the same physical region of sky in all the frames. A `sigma image'
containing the standard deviation of the values corresponding to each
pixel in the final image was also derived.

\begin{figure}[h]
\begin{center}
\includegraphics[scale=0.4, angle=0]{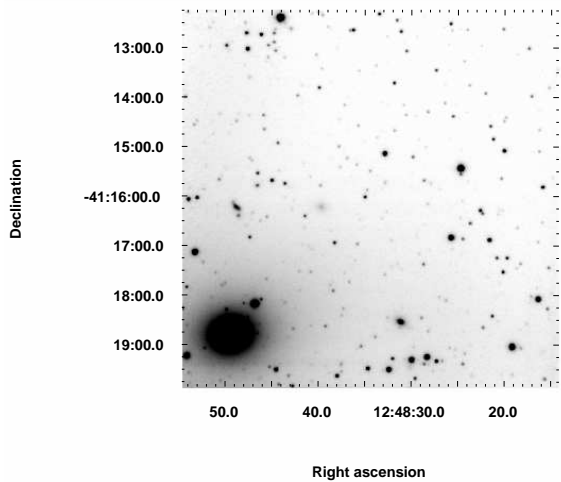}
\caption{Final image used for the analysis. North is to the top and east is
  to the left. The field of view is $7.7\times7.7$\,arcmin. NGC\,4696 is
  located near the south-east (bottom left) corner.}
\label{fig:redimage}
\end{center}
\end{figure}

The final image obtained using this reduction procedure is shown in
Figure~\ref{fig:redimage}. Large-scale (of the order of arcmin)
variations in the resulting background that do not correspond to the
galaxy's light were observed at the level of 0.4\% of the mean sky
background, corresponding to an uncertainty of $\approx 0.004$ mag arcsec$^{-2}$. These variations impose the effective limit on the radius to
which we were able to accurately measure the surface brightness profile.

The photometric zero point for our observations was obtained from the
photometric standards we observed. Our instrumental magnitudes for these
stars were compared to their $H$ magnitude in the 2MASS catalog. The
zero point obtained has an associated error of $\pm0.03$\,mag.

\section{Data Analysis}
\label{sec:an}

Two further elements were needed prior to fitting the galaxy's profile:
an estimate of the PSF in the observations, and a mask for the other
objects in the field. The profile could then be studied using the
two-dimensional fitting algorithm GALFIT \citep{peng02}.


The PSF was obtained from the profile of a bright star observed immediately after NGC\,4696 at a
similar airmass to the science field, and hence with comparable seeing. A region of the image centered on
the PSF star and covering 150$\times$150 pixels ($\approx 67 \times 67$
arcsec) was used directly to model the PSF. This PSF was convolved with
the different galaxy profile models before comparing them with the data
during the fitting process.

In order to test the robustness of the fitting process to the exact PSF, the fits were repeated with artificial PSF images with varying FWHM, and Moffat and Gaussian profiles.  Big differences were found between a Moffat and Gaussian profile, but the FWHM had only a small effect ($\approx 2.5\%$) on the fitted value of $r_{e}$.  Since we used an image of a star for the PSF, the shape of the profile will be accurately determined, and thus our fits should be robust to any changes in the PSF FHWM due to any differences in observing conditions.  Furthermore, the PSF will only strongly affect the core region of the galaxy profile, which was not used in the final fit.


A mask was needed to exclude other objects in the field affecting the
profile fit. For the most of the field, objects were detected using
SExtractor \citep{bertin96}. However objects close to the core of
NGC\,4696 were not correctly detected, so objects in a region out to
$\sim$90\,arcsec from the centre of the galaxy were detected and
masked by eye.


The galaxy surface brightness distribution was fitted using GALFIT. The
models considered were a de~Vaucouleurs profile and the more general
S\'{e}rsic profile. Different initial parameters were used to test the
stability of the fitting algorithm. The `sigma' image obtained from the
combining process was used as the input error assigned to each pixel.

There is a strong dependence of the fit results on the correct
determination of the sky level. We therefore kept the sky level as an
additional free parameter in all the fits. In order to test the dependence on the sky level used, we also fitted profiles with the sky level held constant at $\pm 0.2\%$ of the best fitting value, to reflect the 0.4\% variation seen across the image.  The results are given in Table~1.

The geometric parameters of the galaxy (axis ratio and orientation of
the semi-major axis) were obtained from the initial fits. The axis ratio
is $b/a = 0.83\pm0.01$ and the position angle is $\theta =
100^{\circ}\pm2^{\circ}$ (measured north through east). These values
were kept constant in the following fits, and used to construct masks
limiting the region to fit, setting both the minimum and maximum radius
for each fit.

Masking the core of the galaxy was essential, as some complex structure
was observed in that region that is unrelated to the rest of the
galaxy's profile \citep{crawford05}. The region masked was an ellipse
with axis ratio and orientation as above and semi-major axis of
18\,arcsec.

The goodness-of-fit can be evaluated from the $\chi^2$ of the fit and
the number of degrees of freedom ($\nu$) in the model. The value of
$\chi^2$ is calculated directly by GALFIT. However, due to the
correlation between neighbouring pixels caused by the PSF, $\nu$ can not
be calculated directly as the number of pixels minus the number of
parameters. To overcome this, artificial images were created from models
with the same parameters as the fits obtained and noise added at the
same level as our data. These images were then smoothed using a PSF of
the same FWHM as observed. Fits obtained from the resulting smoothed
images give an estimate of the effect of the correlation. Values of the
reduced $\chi^2$ ($\chi^{2}_{\nu} = \chi^{2}/\nu$) quoted below take
this correction into account.

\section{Results and Discussion}
\label{sec:results}


Successive fits were performed in regions limited by an ellipse with the
same geometric parameters as the galaxy and a varying semi-major axis.
The maximum distance at which the fit was reliable was estimated using
three indicators: the values of $\chi^{2}_{\nu}$, visual inspection of
the model-subtracted image, and the variation of flux with angle in
successive elliptical rings. The maximum allowed value for the
semi-major axis distance was 180\,arcsec
($37\,h^{-1}_{70}\,\mathrm{kpc}$). All the results below refer only to
fits within this region.

\begin{figure}
\begin{center}
\input{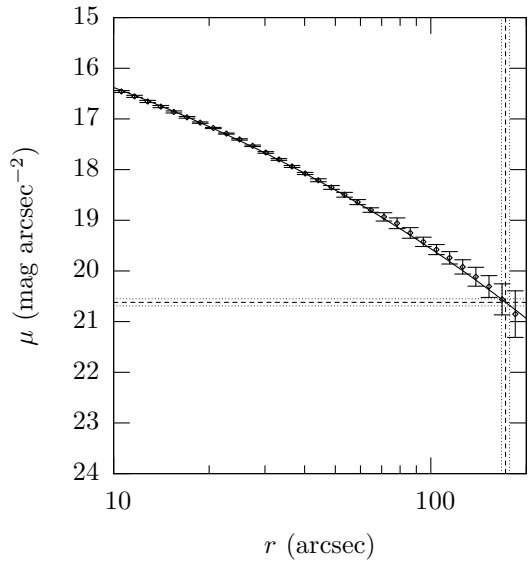}
\caption{Comparison between the observed profile and the fitted
  de~Vaucouleurs model. Each point in the plot corresponds to the
  average surface brightness of the pixels within an elliptical annulus
  (obtained with IRAF task {\tt ellipse}); $r$ is the semi-major axis
  distance to that annulus. The error bars correspond to the dispersion
  in surface brightness of these pixels. The dashed (and dotted) lines
  correspond, respectively, to the values of $r_e$ and $\mu_e$ and their
  uncertainties.}\label{fig:prof}
\end{center}
\end{figure}

The total magnitude ($m$), central surface brightness ($\mu_0$),
effective radius ($r_e$), and surface brightness at $r_e$ ($\mu_e$) 
obtained for de~Vaucouleurs and S\'{e}rsic profiles fitted to the
studied region are shown in Table~\ref{tab:par}. The extra free
parameter in the S\'{e}rsic model (the index $n$) scarcely improves the
value of $\chi^2$, and the index obtained is very close to the
de~Vaucouleurs value of 4. We therefore conclude that this extra free
parameter is not needed, and that the light profile of NGC\,4696 is well
described (in the region within 180\,arcsec) by a de~Vaucouleurs law. A
plot of the observed surface brightness profile of NGC\,4696, together
with this best fit, is shown in Figure~\ref{fig:prof}.

\begin{table*}
\begin{center}
\caption{Surface Brightness Profile Parameters}\label{tab:par}
\begin{tabular}{lccc}
\hline 
Parameter                   & de~Vaucouleurs   & S\'{e}rsic     & Systematic errors$^{a}$ \\
\hline
$m$ (mag)                   & $6.25 \pm 0.06$  & $6.13 \pm 0.10$ & (4.86, 6.94)\\
$\mu_0$ (mag arcsec$^{-2}$) & $12.29 \pm 0.07$ & $11.8 \pm 0.5$ &(8.37, 14.51)\\
$\mu_e$ (mag arcsec$^{-2}$) & $20.62 \pm 0.07$ & $20.90 \pm 0.22$ &(23.75, 19.36)\\
$n$ ~(S\'{e}rsic index)     &  4               & $4.3 \pm 0.3$  &(7.3, 2.4)\\
$r_e$ (arcsec)              &  $172 \pm 5$     & $204 \pm 20$  & (1199, 76)\\
$r_e$ ($h^{-1}_{70}$ kpc)   &  $35.3 \pm 1.0$  & $42 \pm 4$   & (246, 16) \\
$\chi^2_{\nu}$              &  1.170           & 1.167          &\\
\hline
\end{tabular}
\medskip\\
$^{a}${Limits based on fitting a S\'{e}rsic profile with sky values fixed at $\pm 0.2\%$ the best fitting value.}
\end{center}
\end{table*}

The fit was repeated introducing variations of the geometric parameters
as well as the masking of the core region in order to test the
robustness of the results. The main differences were found when varying
the core masking, probably due to the presence of some extra structure
close to the centre of the galaxy or uncertainties in the PSF. Errors
quoted in Table~\ref{tab:par} correspond to these variations. The
differences found when varying the geometric parameters (orientation and
axis ratio) were smaller than these errors.

\begin{figure}
\begin{center}
\input{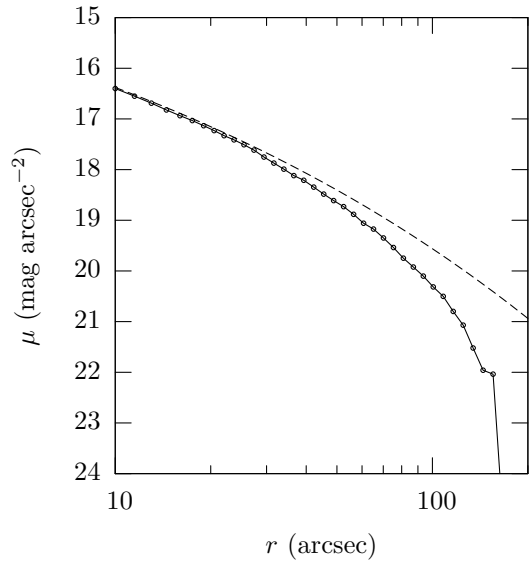}
\caption{Comparison with 2MASS surface brightness profile. The points
  and solid line are the 2MASS profile and the dashed line is our fitted
  de~Vaucouleurs profile (with parameters as in Table~\ref{tab:par}).}
\label{fig:2mass}
\end{center}
\end{figure}

The profile obtained was compared to that obtained in the \Hband from
2MASS. The comparison is shown in Figure~\ref{fig:2mass}. There is a clear difference in the slopes of the 
profiles beyond the galaxy core.  2MASS fit a S\'{e}rsic profile with an index of $n=1.11$, whereas we find a profile that is well fit by an $n=4$ de Vaucouleurs model.  The difference between our result and the 2MASS result is likely to be a consequence of the sky background estimation.   It was found that the slope of the outer profile was very sensitive to the precise sky level used.  If we subtract a sky level higher by only $\approx 0.3 \%$ than that
obtained from the best fit, the difference was essentially eliminated.  Note that this difference could be explained by the large scale variations of $\approx 0.4\%$ observed in our final image, whereas 2MASS has more optimally observed flat-fields.  However, our deeper observations are in better agreement with previous work done in the optical; \citet{schombert87} finds a good fit with a de Vaucouleurs model to $r\approx 5$ arcmin, and \citet{jerjen97} fit a S\'{e}rsic profile with $n=2.04$.  

The extremely  sensitive nature of the fits to the sky level (see Table~1) means that the previous uncertainty in the profile of NGC\,4696 is still unresolved.  Future observations could improve on the present work through either off-source blank field observations interleaved with the galaxy observations, or drift scan observations, as in 2MASS but deeper, to better estimate the flat-field and reduce the uncertainty in the sky levels.

\section{Conclusions}
\label{sec:concl}

The NIR light profile of NGC\,4696 in the \Hband was measured out to a
distance of 180\,arcsec ($37\,h^{-1}_{70}$\,kpc) along the semi-major
axis. This limit was imposed by systematic errors in the flat-fielding
process. The obtained profile is well described by a de~Vaucouleurs law,
with no signs of either truncation or an extended halo. These
observations agree, up to an uncertainty of 0.3\% in the sky level, with
the \Hband profile obtained from 2MASS.

\section*{Acknowledgments}

We thank Tom Jarrett for useful discussions on this work.  This paper is based on data obtained with the Anglo-Australian
Telescope. It also makes use of data products from the Two Micron All
Sky Survey (2MASS), which is a joint project of the University of
Massachusetts and the Infrared Processing and Analysis Center/California
Institute of Technology, funded by the National Aeronautics and Space
Administration and the National Science Foundation.

\end{document}